\documentclass{article}

\usepackage[english]{babel}
\usepackage{eurosym}

\usepackage[usenames]{color}

\begin{document}

\title{Informaticology: combining Computer Science, Data Science, and Fiction Science%
}
\author{%
Jan A. Bergstra\\
{\small  Section Theory of Computer Science,
Informatics Institute,} \\
{\small Faculty of Science, University of Amsterdam,}\\
{\small Amsterdam, The Netherlands.}%
\thanks{Author's email address: {\tt j.a.bergstra@uva.nl}. 
}
\date{}
}

\maketitle

\begin{abstract} 
\noindent Motivated by an intention to remedy current complications with Dutch terminology concerning informatics, the term 
informaticology is positioned to denote an academic counterpart of informatics where informatics is
conceived of as a container for a coherent family of practical disciplines ranging from computer engineering and software engineering to network
technology, data center management, information technology, and information management in a broad sense.

Informaticology  escapes from the limitations of instrumental objectives and the perspective of 
usage that both restrict the scope of informatics.
That is achieved by including fiction science in informaticology and by ranking fiction science on equal terms with computer science and data science, and framing (the study of) game design, development, assessment and distribution, ranging from serious gaming to entertainment gaming, as a chapter of fiction science. A suggestion for the scope of fiction science  is specified in some detail.

In order to illustrate the coherence of informaticology thus conceived, a potential application of fiction to the ontology of instruction sequences and to software quality assessment is sketched, thereby highlighting a possible role of fiction (science) within informaticology 
but outside gaming.
 \end{abstract}

\section{Introduction}\label{sec:Intro}
The terminology about and around computer science constitutes 
a topic that produces puzzling questions for an employee of a 
computer science related department of a Dutch University. The Netherlands probably don't possess the size required for a sustainable 
development of a full range of terminology, in Dutch language, for a field that mainly exists and originates outside the Netherlands. Therefore in areas featuring a multitude of non-Dutch terms a  delicate balance needs to be found between local and global conventions regarding the design of naming schemes and terminology. 
 
This balance is difficult to find if the conventions in the source languages of the non-Dutch terms are steadily evolving. Such an evolution takes place in and around information technology.%
\footnote{For instance:  grid has been surpassed by cloud in just a few years time; supercomputing has become high performance computing and outsourcing strategy has become sourcing strategy; personal computers have become desktops and cell phones have become smartphones.}

Designing alternative terminologies is an option for dealing with the co-evolution of international and local jargon. Below I will develop some new options for naming conventions and I will contemplate a range of direct and indirect consequences of these proposals.
 
\subsection{Objectives of the paper}
I intend to deal with three issues: (i) complications with Dutch terminology (in Dutch) concerning informatics and related matters, and a proposal for a novel terminology with as a side effect the need to make use of the somewhat unusual term informaticology in English; (ii) A definition of the subject matter of informaticology with as a novel entry ``fiction science", which is unusual to the extent that providing an attempt to  include a definition of it is justified, as well as a rationale for having it included it in informaticology; (iii) An attempt to illustrate the coherence of informaticology thus obtained by pointing out how fiction 
may play a role in quality assessment methods for system engineering.

\subsection{Survey of the paper}
In Section \ref{Isvp} various issues concerning Dutch terminology are discussed and then informaticology is characterized  as the sum of three different but overlapping components, or rather the topic space generated by these three components: computer science, data science, and fiction science. 

Fiction science being an unusual item in this list, in Section \ref{FSexplained}
an extensive justification and explanation of the phrase ``fiction science" is given. Then in Section \ref{CSS} the connection with complex systems science is discussed, which is considered to lie outside informaticology but which at the same time can be fruitfully combined with informaticology. In Section \ref{FSinseq} possible roles of fiction (science) in the theory of instruction sequences are described culminating in the suggestion made in Section \ref{TIUD} that in some cases, fiction is an indispensable for making sense of quality assessment methods. Some concluding remarks are made in Section \ref{Conclusions}.

\section{Informatics: science versus practice}
\label{Isvp}
The connotations of the term informatics depend on the language from which it is considered. I will look at the term from the perspective of the Dutch language  because there I see a need to propose some new terminology. Incentives for these proposals stem from my experience in departmental organization.

I will now formulate three competing assumptions on how informatics might be translated into Dutch. Of these options I favor assumption A1, mainly because in my view it leads to the best design of Dutch jargon about informatics. Admittedly the match with internationally accepted terminology is as yet rather poor.

\subsection{Assumption A1: EN2NL(informatics) = {\em informatica}}
To begin with, I will assume that the Dutch translation of informatics is {\em informatica}. 
This assumption, however plausible it may seem at first sight, should not be taken for granted.

The term {\em informatica} in Dutch seems to have the following connotations: (i) somewhat narrow-minded
and limited to dealing with computers in conventional ways, (ii) reflecting an old-fashioned approach to information technology research, (iii) representing a failed attempt to introduce a new term inspired by existing (Dutch) terms such as {\em fysica} (physics), {\em genetica} (genetics), {\em dynamica} (dynamics), {\em hydraulica} (hydraulics), (iv) an orientation towards academic competence rather than towards professional and practical competence. Informatics seems not to have these unfortunate  connotations, except that it seems to share with {\em informatica} an academic and scholarly bias.

Unfortunately these narrow connotations of {\em informatica} render the term somewhat useless in The Netherlands. Either its meaning ought to be changed, which seems to be quite difficult, or new terminology needs to be introduced. In this paper I have chosen to contemplate the second approach. I propose to use {\em informaticologie} as a term to indicate an academic field of expertise (a science) related to {\em informatica}.  {\em Informaticologie} may replace many uses of the term {\em informatica}.  Moreover I will assume that informaticology is an adequate English translation of {\em informaticologie}.%
\footnote{One might object that
{\em informatiekunde} (information science) already occupies the (conceptual) niche that  
{\em informaticologie} is supposed to inhabit. Unfortunately information science has complicating connotations rendering it of limited use for instance the mathematical bias of information theory, while {\em informatiekunde} has a definite bias towards business informatics (in Dutch). Some claim that {\em information sciences} 
(information sciences), the plural constituting an essential feature of the phrase, can (and should) be used in Dutch for denoting a very general academic counterpart of {\em informatica}. The latter view might prevail, though I prefer the use of {\em informaticologie} instead.}

{\em Informaticologie} does not replace {\em informatica}. On the contrary, I propose that instead of 
denoting a mainly academic discipline, 
{\em informatica} should denote (in Dutch) a range of practical activities and competences,  
a practice rather than a science so to say. Correspondingly an {\em informaticus} typically intends to achieve practical objectives, in a way comparable to a {\em musicus} (musician), and a {\em politicus} (politician). The academic areas from which the bodies of knowledge for respectively the {\em musicus} and the{\em politicus} are derived are {\em musicologie} (musicology) and {\em politicologie} (political science). I propose to use {\em informaticologie} 
(informaticology) for the 
academic background of an {\em informaticus} (assuming he or she has one).%
\footnote{For the English terminology these choices imply that informaticology ought to denote (or to evolve in such a way that it will denote) the science of informatics if a best fit with the proposal on Dutch terminology is to be found. I have no opinion concerning the plausibility that the meaning of the terms informatics and informaticology will evolve in that direction. My proposal concerning Dutch terminology is largely independent of the current and future semantic relation between informatics and informaticology.}

An individual knowledgeable of {\em musicologie} is called a {\em musicoloog}, and a scholar of {\em politicologie} is called a {\em politicoloog}. This suggest that an academically oriented expert of {\em informaticologie} may be called an {\em informaticoloog}.%
\footnote{There are other examples in Dutch that don't support the case that I intend to make: {\em psychologie} (psychology) covers both an academic discipline and its practical counterpart at the same time, a {\em psycholoog} may be either an academic or a practitioner.} 
Translating {\em informaticoloog} to informaticologist is perhaps plausible but expecting that
informaticologist will become a successful phrase in English is probably unwarranted. 
I see no objection, however, against using the term 
{\em informaticoloog} in Dutch, apart from the obvious difficulty that it is quite unusual at this moment.

Summing up, I favor informaticology%
\footnote{In January 2012 writing as  @\texttt{janaldertb} I have 
posted 5 tweets about ``informaticology'' including
the following translations between English (EN) and Dutch (NL): EN informaticology = NL informaticologie; EN informaticologist = EN informaticist = NL informaticoloog, 
EN informatician= NL informaticus = NL informaticien; EN theoretical informaticologist = NL theoretisch informaticus.
In Folia Magazine (\texttt{http://www.foliaweb.nl/organisatie/directeur-ivi-pleit-voor-brede-faculteit- informaticologie}) of the first week of February 2012 these terms were explained together with their possible application in the University of Amsterdam.}
over informatics for denoting an academic discipline%
\footnote{Consistent with this preference I would currently prefer my own affiliation at the University of Amsterdam to have been called ``Institute of Informaticology''  rather than its current name  ``Informatics Institute". I don't claim that the significant cost of a name change would be compensated for by that impovement, neither do I claim that for all, or even some, other staff members of our Informatics Institute the same preference would hold.}
because in The Netherlands (and therefore in Dutch language)  ``informatics'', that is to say its Dutch translation {\em informatica}, has unfortunately acquired a connotation of narrow-minded (low) tech which is hard to defeat irrespective of the success of ``informatics'' as a term outside The Netherlands.%
\footnote{For instance after studying {\em Informatik} at a University in Germany a student has become an {\em Informatiker} ({\em Diplominformatiker}). That compares well with {\em Mathematik} and {\em Mathematiker} ({\em Diplommathematiker}). In Dutch, however, the counterpart {\em informaticus} of the German {\em Informatiker} has never become fashionable and {\em IT-specialist} (IT specialist) has become the dominant label of a professional with a higher education background in {\em informatica}. Now unfortunately everyone employed (in The Netherlands) in the IT sector availing of a higher education background in whatever field is labeled an {\em IT-specialist} and equally unfortunately highlighting that someone has obtained an MSc in informatics (which is formally permitted) is still very unusual.}

\subsection{Assumption A2: EN2NL(informatics) = {\em informatics}}
The line of reasoning based on assumption A1 becomes unconvincing if one prefers to translate informatics to {\em informatietechnologie} (information technology), to {\em computer science} (computer science) which is often used in Dutch%
\footnote{Dutch is quite flexible in adopting English words and phrases literally as parts of the (Dutch) language ready for daily use and preferred over somewhat artificial (or seemingly out of date) Dutch translations. Here are some examples: penalty, goal (in soccer), game, break, gamepoint, machtpoint (tennis), computer, laptop, software, printer, beamer (computer based electronic equipment), overhead projector, assessment, bachelor, master (education), hit (pop music), life science (academic expertise area).}
by lack of a better alternative phrase in native Dutch, or to {\em praktsiche informatica} (practical informatics). Remarkably the problem with {\em computer science} (now understood as a phrase ready for use in Dutch just as the very successful imported phrase {\em life science}) is not its plain non-Dutch origin, which goes by almost unnoticed for a Dutch speaking audience. Problematic is the hardware bias that comes along with ``computer''. The distance between computer and computing is very large in Dutch.%
\footnote{Computing has no common translation in Dutch. The closest term is {\em berekenen} but that has a very strong bias towards arithmetical computation in an applied mathematical context. Some people may even translate computing to {\em processing}, again a term for which no convincing Dutch translation has been found. The closest to a translation of processing comes {\em verwerken} which has a strong non-mathematical bias, however, and in fact rather has an administrative connotation. An old suggestion promoted by E.W. Dijkstra is to use {\em computing science} instead of {\em computer science} but that has never worked out and I see no hope of recovery for the phrase {\em computing science}, at least not  in The Netherlands. 
Due to historic path dependence that phrase has become co-extensive with a quasi formal approach to software engineering. Currently {\em computational science} (computational science) is becoming common in Dutch IT speak. It represents the computational approach to science rather than the science of computational phenomena.}

Each of these translations has its disadvantages. Perhaps an efficient terminology results if one translates informatics by {\em informatics} (that is assumption A2 on EN2NL(informatics)) thereby importing informatics as a novel Dutch term which is (supposed to be) free from the unfortunate connotations that {\em informatica} has acquired in the last five decades. After adopting A2 and replacing {\em informatica} by {\em informatics}, however, the design of Dutch terminology will leave issues unsettled to an extent which makes me favor to work from assumption A1 above instead of adopting A2.

\subsection{Assumption A3: EN2NL(informatics) = {\em informaticologie}}
Assumption A3 claims that informatics is best translated as {\em informaticology}. Assumption A3 might be an option worth of further contemplation. It would require that {\em informatica} is translated to a term or phrase with a practical connotation, for instance information technology, practical informatics, or applied informatics, or 
alternatively that {\em informatica} is left without a translation and is decomposed into a range of components such as {\em bedrijfsinformatiekunde} (business informatics), {\em medische informatiekunde} (medical informatics) etc. each of which seem to have unproblematic translations in both directions.

\subsection{Informaticology decomposed: IY = CS + DS + FS}\label{informaticology}
Informaticology (IY) contains the following three compartments, each consisting of a large variety of specialties.
\begin{description}
\item{\em Computer Science (CS).} Computer engineering and technology, quantum computing, processor architectures, multi-core processing, concurrent systems engineering, operating systems, software engineering, computability theory, robotics, embedded and real time systems, algorithms, cryptography, human-computer interaction, computer security, anonymity (from the systems perspective), system and software validation, IT management, 
\item{\em Data Science (DS).} Grid, cloud, big data, data visualization, high speed networking, information systems, information theory, information management, information science,  ontology (ontologies), anonymity (from the data perspective), machine learning, e-science.
\item{\em Fiction Science (FS).} Game technology, game design (both serious and entertainment), interactive poetry, interactive literature, interactive fiction, interactive fancy (fiction outside the logically possible), and parts of e-humanities.
\end{description}

These three parts have overlaps and intermediaries of various forms, finding an othogonal decomposition informaticology was not intended. Writing P(H) for ``H in practice" and INF for informatics we find INF = P(IY), and after elimination of IY: INF = P(CS + DS + FS).

The ``sum" CS + DS + FS should be understood as combining generators rather than as denoting a disjoint union or a near disjoint union. For instance machine learning, high performance computing, machine translation of natural languages, and the non-brain science/non-psychology part of cognitive science each emerge in the cone of subjects generated by CS and DS.

\subsection{{\em Informaticologie} needs ``informaticology''}
One may say that my objective to deposit a 
proposal to use {\em informaticologie} in Dutch instead of {\em informatica} in 
Dutch merely justifies  
writing a Dutch text and posting it somewhere locally in The Netherlands, rather than to write a text which is arxived internationally. As an argument against that position I propose to acknowledge that the dependence of Dutch on English, and the dependence of The Netherlands on an English speaking international community, has become so strong that Dutch jargon unequipped with convincing (in the eyes/ears of a Dutch audience) translations is deeply distrusted by Dutch native speakers.

Thus once {\em informaticologie} has been coined and approved of (as I have done above) it must be provided with an English translation, even if that translation is disappointingly uncommon, otherwise its Dutch usage is blocked at once.

\subsection{Implications for {\em informatica} curriculum design}
Most Dutch universities offer a bachelor degree in {\em informatica}. One might ask if the proposed naming scheme implies that such degrees should be renamed into  {\em informaticologie}. This might be advisable (assuming the suggested naming convention) provided the curriculum contents have a focus on scholarly investigation and research. 

Much more effective might be to have the focus of an {\em informatica} curriculum manifestly directed towards practice. This also requires a strategic change. Nowadays (in The Netherlands), mainly the medical schools and the law schools provide university degrees for what is in essence a vocational training. Informatics (IT) has become so complex, however, and the modern world has become so dependent on it,  that it is definitely not to be considered  a waste of talent if someone chooses to follow a career based on technical competence in that area. Higher education should support such career paths by offering strong vocational preparation properly embedded in corresponding degrees. 

An important aspect to this matter is that nowadays industry seems to be in de lead of innovation in informatics rather than academic research. Innovation in informatics (practically conceived) is a must rather than an option for a modern society. Now separating research from innovation may be artificial in informatics, and promoting innovation outside the scope of industrial objectives may be an uphill struggle. I conclude that universities must commit themselves to the education of a workforce which can carry out the many complex tasks in informatics that lie ahead. The viewpoint that such work will be intellectually less demanding than research and scholarly work is flawed. The viewpoint that other educational schemes will produce a workforce which propels the field while academic graduates lead the path of development and innovation from their lofty research institutes is outdated.

Offering university degrees at both bachelor and master levels in well-known practical subareas of informatics is currently uncommon in The Netherlands (specially at the bachelor level, where wide spectrum curriculum design flourishes for local reasons). But in my view that will prove to be very effective. Titles might for a bachelor degree may include: software engineering, system and network engineering, computer engineering, information retrieval and big data technology, game design and technology.

Of course offering a research (rather than innovation) oriented degree in {\em informaticologie}  is an option too, but that is no substitute for offering vocational training at the highest intellectual level to a significant number of students in informatics.

\subsection{Informaticology: futile isolationism?}
One may object to any attempt to provide a profile around computer science, informatics, data science etc. on the grounds that future work needs to take a much wider perspective into account, a perspective 
nowadays still termed multidisciplinary (a
phrase that seems to indicate implicitly that the envisaged state of affairs has not yet been reached). That wider perspective must   include nanotechnology, solid state physics, biotechnology, genetics, geography, demography, social sciences,  medical sciences, finance, philosophy, (divinity?), and so on. 

If the wider perspective is seen as both essential and critically absent in the current phase of development of ``the field",  then working towards a demarcation  of informaticology as a moderately widened container of academic informatics (``informatics as a science") may be considered a sign of intellectual provincialism, aimed at a futile isolation of a field that should not even ``try" to have an independent existence. As a reply I would point at law, which is a field that profits from an independent existence while being connected to and dependent of issues outside law by definition.

The competence of a lawyer is comparable to the competence of an informatician in that both will always 
need to be instrumental in serving objectives originating from outside law and informatics respectively.
Multidisciplinary approaches must not be based on the prejudice that a disciplinary approach cannot be open minded. Open mindedness must not be promoted as a substitute for a multidisciplinary approach. Who talks about the virtues of a multidisciplinary framework always subscribes to the fundamental role of mono-disciplinary work in the first place. Therefore even the strongest advocates of a multidisciplinary approach in which computer science, data science, etc. feature as signifiant components need to worry, or at least be aware, of the ``identity" of the single (or mono) discipline from which such competences are taken.

\section{Fiction Science composed and justified}
\label{FSexplained}
The component of informaticology, as proposed in paragraph \ref{informaticology} above, that is most in need of motivation and explication is fiction science. The phrase ``fiction science'' understood as ``the science (and technology) of fiction'' seems not to be in regular use. Nevertheless I see ample justification for placing it at the same level as both other categories. 

\subsection{Fiction science: scope and size}
Here is  list of reasons and justifications for taking the phrase fiction science seriously.
\begin{enumerate}
\item Computer games for entertainment have come to stay.
\item The power of computer games far exceeds the existence of successful programs for playing chess or checkers,
or the availability of highly realistic flight simulators for a range of past, present, and perhaps even future aircraft. The 
fundamental strength of computer gaming lies in the ability to capture people's imagination by presenting them individually and collectively an alternative existences and with unforeseen challenges within artificial worlds 
\item Serous games have come to stay and  will gradually become much more important. Serious games in teaching  will approach the importance of current school books (and perhaps even schools) within a few decades.
\item Computer gaming is about much more than advanced graphics, agent theory, simulation, and applied artificial intelligence.
For computer gaming the systematic usage of fiction is a critical factor. Fiction seems not to have a
counterpart in either computer science or in data science.
\item Due to computing fiction will become interactive. The ability of computer gaming to bring life into fiction is an essential strength and an explanation of its major contribution.
\item If gaming is used as a top-level category, then it will figure at the same level as computer science and data science, a state of affairs which is rather implausible. So another top-level category must be found that may stand on equal footing with computer science and data science. Gaming, both serious and for entertainment, need to be captured in an overall category independent from computer science and data science and the proposal made here is that ``fiction science'' may be the name of that category. This proposal is made in spite of the fact that the phrase 
``fiction science" seems not to have been in regular use for that purpose and with that kind of meaning already.%
\footnote{This proposal  is definitely not meant as a pun or as an alternative way to refer to science fiction.}
\item Fiction science (science of fiction) encompasses the use of science fiction (fictional science) as a driver for fiction design. Fiction science cannot provide a scientific basis for science fiction, but it can help to achieve predetermined goals by means of interactive fiction that is based on story lines inspired by science fiction. Science fiction research (that is research about science fiction, if performed at all) needs to be placed in an overarching category just as gaming research and fiction science seems to be adequate for that role.
\item Computer technology can transmit fiction at least as good if not better than fact. This may be an understatement: computer technology opens the door to interactive fiction, which is perhaps causing a more disruptive in the field of fiction presentation than interactive data causes in the field of (factual) data presentation.
\item Some new fictional characters will become more widespread and influential than 
fictional characters have been till today.%
\footnote{It is commonplace to claim that Arthur Conan Doyle's fictional character and London detective Sherlock Holmes is more well-known than any currently active (real) detective anywhere in the world. The same may hold for the German fictional detective Derrick, made famous on German TV by Horst Tappert.%
}
\item Fictional characters of the future may outperform even the best known real characters of the time.
\item The relative (cultural) weight of fiction will increase at the cost of fact.
\item For decision taking purposes the application of scenario development and analysis will be needed increasingly. 
A scenario in most cases deviates from reality in ways comparable to the deviations of physical laws. Both scenario's and fiction have common philosophical roots. Scenario design and analysis technology  builds on computer fiction.
\item Fiction science has an attractive philosophical basis, which escapes from the narrow realist ontologies which have proven to be very fruitful in the natural sciences as well as in computer science and in data science.
\item It is my expectation that the future growth of ``computer based fiction''  (resulting from fiction engineering) will be phenomenal. Fiction may come to stand in between (natural, social, computer, and data) science and religion, and at par with sports and politics, and we won't soon find out which of these forces is the stronger one. Fiction may at some stage outperform (empirical and science based) fact as the primary content of computer based data storage and computer mediated communication, because of its superiority  in capturing the attention of a large audience.
\item Fiction science is needed to provide the production and rendering of fictional information with the backing of an experience base which has been reflected upon critically and systematically.
\end{enumerate}

\subsection{On the semantics of fiction}
Much has been written about the meaning of fictional accounts. I mention some informative papers: 
\cite{Everett2007,Lamarque1990,Proudfoot2006}. A recent survey concerning the logic and philosophy of fictional episodes and fictional objects can be found in \cite{Muller2012}. Important for Section \ref{TIUD} below is the distinction between fictional stories that can be told as if they were reports on pretended fact, hypothetically held true in some possible world, and fictional accounts that defeat a reasonable possible world interpretation (sometimes referred to as fancy rather than as fiction). In Section \ref{TIUD} I will make use of non-fancy fiction. The semantic challenges posed by fancy seem to be greater than those posed by mere non-fancy fiction.

The formidable literature on the semantics of fiction suggests that much more work needs to be done before a final picture emerges. I cannot resist formulation at this place some remarks on the matter pertaining to non-fancy only, however.
\begin{enumerate}
\item In mathematics arguments are often conducted by working towards a contradiction from an assumption $\phi$, (on top of assumptions $\Phi$) then to conclude the negation of $\phi$. Reasoning in the combined system $\Phi$+$\phi$ 
is more speculative than most fiction because such a world will turn out to be non-existent in a very strong sense. It is impossible that one fully imagines such a world because it cannot exist. Rather one may merely  apply formalistic reasoning to it.
\item A fictional story has a form of existence that can be compared with a dedicated mathematical structure meant for modeling a specific application area. Perhaps one may imagine the universe of all possible fictional stories as a class of structures (comparable to the processes in process algebra \cite{BaetenBastenReniers2009} which exist prior to their being specified) existing already inside a mathematical universe which itself is embedded inside the cumulative hierarchy of sets. In that view dedicated mathematical models only specify (thereby rendering them more accessible for human thought) structures that were already present in models of set theory (that is present as sets) before their usage was contemplated by an applied mathematician (or computer scientist), fictional stories exist before being produced. 
\item The complication that fictional stories may become logically inconsistent and thereby specify fictional worlds that cannot exist after all, should not be overrated. Set theory in mathematics is a particular way of removing inconsistencies applied in a setting where inconsistencies can not be excluded in advance. Thus if the risk that fictional stories are inconsistent is used to deny all (objects occurring in) fictional stories some abstract existence, transposing that that line of thinking to mathematics would rule out Zermel-Fraenkel set theory because of the proven risk that a plausible modification of set theory (of Cantor's original version) is inconsistent (Russel's paradox). Assuming that most mathematical objects exist only under the assumption of consistency of a theory about them, it is plausible that entities within appropriate fictional stories can exist in a similar way, under the condition that the fictional stories are not too inconsistent.
\item If fiction leads to circumstances where logical contradiction is unavoidable, one may contemplate the  introduction of a fictional logic that governs reasoning inside the fictional narrative. This is not unlike working with paraconsistent logics (see \cite{Middelburg2011a}).
\end{enumerate}

\section{Complex systems science: outside informaticology}
\label{CSS}
Classification of informatics related themes in the three subcategories of informaticology may be problematic in the sense that some themes overlap with say both CS and DS (e.g. very large data bases), or with both CS and FS (e.g. graphical techniques for rendering games). This indicates that CS, DS, and FS merely provide a focus, while not indicating technical disjointness. Some themes, however, might fit within informaticology at first sight but will be excluded at closer inspection. For some themes one may disagree about a proper classification. 
I would classify visual computer art in DS, and machine translation of natural languages in CS. Software patents and software copyrights in CS,  but one might argue that these are topics in law that lie outside the scope of informatcology. The societal impact of social media, however, definitely lies outside informaticology, because that is better viewed as social psychology or sociology.

The study of complex systems, also called ``complex systems science" (CSS), pursued from a multi-disciplinary perspective has become a fruitful and important perspective. It may be tempting to classify complexity science under informaticology as well. I have chosen not to do so because that would create an unrealistic import of different natural and social sciences into informaticology. Now complexity science and informaticology can very well be studied simultaneously and to their mutual benefit  in the same unit, and an academic institution having among its units or departments an ``Institute of Informaticology and Complexity Science'' seems to be a plausible and even attractive state of affairs. There seems not to be a need for a category that integrates complex systems science with informaticology. Cybernetics (CYB) traditionally covers that area, that is the combination of IY and CSS  except FS.%
\footnote{Thus: CYB + FS = IY + CSS.}

\section{Fiction science and instruction sequence ontology}
\label{FSinseq}
This section contains a personal perspective on the aspect of the proposed decomposition of informaticology that I expect readers to consider the least convincing one, that is the listing of fiction science on an equal footing with computer science and data science. 

In this section I will mention a line of thought which led me to contemplate fiction science 
from the standpoint of conventional computer science, and in particular from the perspective of an approach to the theory of software engineering. In the next Section these considerations will be made more concrete in the context of software quality assurance.

\subsection{Ontological issues concerning instruction sequences}
In our recent \cite{BergstraMiddelburg2012b} we report on the results of over 10 years work on instruction sequences and thread algebra (\cite{BergstraMiddelburg2007}) viewed as a topic in computer science.%
\footnote{In \cite{Bergstra2012c} a detailed argument is provided concerning how (and why) a theory of instruction sequences may be used as a model of a theory of computer programs.}
In spite of its conceptual simplicity, if not outright naivety,  working out a philosophy of that subject presents several serious complications. For instance one may have worries about the nature of instruction sequences (for which I will use the abbreviation inseqs with singular inseq), that is their place in an overarching ontology. Are  inseqs mathematical entities and are they to be considered abstract objects for that reason? Or is an inseq a physical object existing in space and time while residing on a suitable information carrier, comparable to the way a painting may reside on a wall
of a building or on a piece of wood, or a surface made of some fabric. Is theory of inseqs about the former (abstract entities) and is the practice of inseqs about the latter? 

Answering such questions is primarily difficult because the philosophical foundations of ontology can be worked out in different directions. Approaches to the principles of ontology range from so-called extreme mathematical realism (all physical objects are reducible to mathematical ones, see \cite{Tegmark2008})  to extreme nominalism (all mathematical objects are reducible to (bio)physical processes in the nerve systems of living agents, see \cite{Feng2007}). Each specific perspective on the principles of ontology leads to its corresponding and equally specific position concerning the status abstract versus concrete entities in general and 
concerning abstract inseqs versus concrete inseqs in particular.

An ontology of programs has been studied in \cite{EdenTurner2007}. That work fails to provide is not what I expect of an ontology of inseqs, and it gives no clue concerning what a program might be either. The explanations of \cite{EdenTurner2007} seem to assume that one knows already what is meant with ``computer'', ``program", and ``execution", a point of view that I do not share. Instruction sequence theory must also explain such notions.%
\footnote{Using the approach to a formal ontology of artefacts in \cite{Kassel2010} progress in the direction of defining what programs are  may be found. A quite comprehensive approach to the ontology of requirements can be found in \cite{JuretaMF2009}.}

\subsection{Abstract objects versus concrete objects}
It is not uncommon to assume that concrete objects exist in space and time while abstract entities do not.%
\footnote{General remarks on abstract objects can be found in \cite{Rosen2012}. A systematic survey of abstract and concrete entities  is found in \cite{Hirst1991}.} 

In \cite{Yablo2002} the essential difference between abstract objects and concrete ones is identified in the fact that abstract objects are characterized by their essence, while concrete objects are characterized by their accense, that is the collection of their accidental properties (qualia in the terminology of \cite{JuretaMF2009}).

Further abstract objects may be found by abstracting away inessential aspects from concrete objects. Abstracting is a difficult matter: one may not be able to forget about the history of an object even if one intends to do so.%
\footnote{Anonymity as a topic within security results from that difficulty. It is meaningless to say that certain information will be forgotten if there is no convincing explanation regarding how that is actually done.} Abstract objects then may appear as idealized limits resulting from a successive chain of abstraction steps. The abstract object then becomes an inverse limit of a sequence (potentially of unbounded length) of successively simplified concrete objects. 

Like all mathematical objects featuring in so-called minimal algebras,  inseqs are generated by means of constructors 
(often called primitive) from basic (or primitive) elements (often called generators). Generation is a hypothetical production process that does not involve a human operator, designer,  or maker. The primitives of an algebra of inseqs  generate an infinity of inseqs at once so to speak.

\subsection{Fictional objects and divine objects}
When making a distinction between abstract and concrete entities, it is meaningful as well to refer to the possible existence of fictional objects as well as to the possible existence of divine objects, thus obtaining four very different ontological classes. Unlike fictional objects, as will be noted below, divine objects (see \cite{Leftow1990}) are quite remote from instruction sequences and from any artifact connected with software engineering, but two remarks are in place (i) divine objects share with abstract objects their essential simplicity, and (ii) it is in relation to causation that abstract objects and divine entities seem to differ most significantly: (the existence of ) abstract objects may be caused by other entities, but abstract entities will not serve as causes for either events or for the existence of other entities. Thus while divine entities may serve as causes of endurant phenomena (existence of entities), or of perdurant phenomena (events, traces), divine entities 
are supposed not to have been caused by any other entity or event (enduring or perduring phenomenon).%
\footnote{Thus an inseq viewed as an abstract entity cannot cause a  machine to display certain behavior, while 
viewed as a divine entity (however implausible that may be) it might have the desired causal effect,}
From some naive perspective one might hold the following order of decreasing plausibility (that is requiring a decreasing ontological commitment) of existence: concrete, abstract, fiction (non-fancy), divine, fancy.

Assuming that a fictional inseqware engineer $E$ has produced an inseq which provides a valid solution to the so-called Halting problem (see \cite{BergstraMiddelburg2012a} for a detailed account of inseqs and the corresponding halting problem for inseqs), leads to fancy rather than to fiction, unless one accepts that ``solving the halting problem" can be a fictional quality (or rather a quale in the terminology of \cite{JuretaMF2009}, that is a subrange of a quality space consisting of a collection of atomic quales) of some inseqware while not being true of it.

\subsection{Inseqware: concrete manifestations of inseqs}
I propose that {\em inseqware}%
\footnote{The primary justification of the work on inseqs consists of that work serving as a model for theories of programs simplified by ``inseq" being simpler to define than ``program". Now inseqware is supposed to stand to inseqs as software stands to programs. Understanding software as ``stuff made out of inseqs''  is not an adequate view because of the (intended) technical limitations imposed on inseqs, and for that reason a new term is needed to denote ``stuff made out of inseqs".} 
refers to physically represented units of information that embody instruction sequences or families of polyadic instruction sequences. This allows to use the convention that the phrase instruction sequence refers to a mathematical entity. Now an inseqware engineer produces inseqware, (while inseq engineers  don't exist by definition). 

\subsection{Fictional inseqware: between abstract and concrete}
Besides  a significant number of papers on abstract versus concrete objects there exists an even more extensive about the ontology of fictional objects. That literature seems to be more to the point for the issue at hand than the ontological questions about abstract versus concrete which are mostly inspired by worries concerning the foundations of mathematics. In particular I hypothesize that a convincing account of the ontological status of instruction sequences (abstract inseqware) and (concrete) inseqware will be found more easily if in addition fictional inseqware is being analyzed as well.

Fictional inseqware results when a fictional inseqware engineer, say $E^{iw}_f$, carries out a (fictional) inseqware development process and delivers result $R^{iw}_f$. Every inseqware engineer who has not been identified unambiguously in person is fictional. By consequence fictional engineers abound in the software engineering literature and when their outputs have not been presented in all detail, which then would permit an abstract view, such results must be categorized as being fictional too.

\section{Taking inseqware usage decisions}
\label{TIUD}
Usage of inseqware requires a machine (execution architecture, see \cite{BergstraPonse2007a}) which puts it into effect (see \cite{Bergstra2011a,Bergstra2012a}) in a context where intended objectives can be obtained (otherwise the effectuation is at best merely a test, see also \cite{Bergstra2012c, Middelburg2010b}). I will speak of actual usage of $W$ when $W$ is effectuated in a context where the intended objectives that $W$'s designers
(or owners) had in mind count.%
\footnote{This makes usage dependent of a designer, or of an owner, which may not be considered convincing.}

Once inseqware $W$ has been produced its usage must sometimes be preceded by taking a decision to that extent (see \cite{Bergstra2012b} for an explanation of my preferred usage of the phrase ``decision taking") that such usage is permitted. A positive usage decision will coincide with granting per mission to use. After the persmission to use $W$ has been obtained both manual and automated actions may trigger the effectuation of $W$. Such effectuations may take place long after the usage permission was granted. In particular for safety critical inseqware actual usage and the option of forthcoming usage under certain conditions can hardly be distinguished because both need to be taking into account when making a proper risk assessment. 

Nevertheless a failure of $W$ can occur only during actual usage and the it necessarily occurs as the consequence of a fault in $W$. A risk of failure, however, is present independently of actual usage as long as the the option of (forthcoming) usage exists, that is the probability of actual usage is non-zero. Thus, the risk of (current or future) failure of $W$ arises as a consequence of a usage decision whether that decision triggers immediate usage or usage at a predefined moment in time or not.

Taking a decision in the style of \cite{Bergstra2012a,Bergstra2012b} may be considered too cumbersome in less demanding circumstances. A weaker step that may mark the transition from design, development, and production phases of $W$ to  its phase of usage is the choice, (or action) to accept (and perhaps even to promote) usage.

\subsection{Usage is actual usage plus contingent usage}
As has been indicated above, actual usage must be distinguished from  forthcoming or potentially forthcoming usage, both of  which I will refer to as contingent usage. Thus contingent usage  is usage that may take place in a later stage (then constituting actual usage). Whether contingent usage of $W\!$ turns into forthcoming actual usage depends on the future evolution of the system $\!$ of which $W\!$ is a component. 

Usage consists of a mix of actual usage and contingent usage. Contingent usage may be considered identical to a non-zero probability of current or future actual usage.  Contingent usage can only take place after either a usage decision is taken, or the choice is made that usage will be acceptable. 

Inseqware $W\!$ is safety critical as a component of a system $S\!$ if there is a non-zero probability of catastrophic system failure caused by a failure occurring during effectuation (that is actual usage) of $W\!$. Because this definition is  difficult to apply in practice the following definition is preferable: Inseqware $W\!$ is safety critical as a component of a system $S\!$ if the risk that there exists a non-zero probability of catastrophic system failure caused by a failure occurring during (and as a consequence of) effectuation of $W\!$ cannot (yet) be excluded.%
\footnote{An unpleasant consequence of this second definition of being safety critical is that simply by acquiring a better understanding, leading to a better risk analysis, of a system a component of it can cease to be safety critical for it.}

When contingent usage of $W\!$ turns into its actual usage, the effectuation of $W\!$ takes place under specific conditions, which will be referred to loosely as parameter conditions, or parameter settings. It will usually depend on the parameter conditions if during the corresponding actual usage a particular fault occurring in $W\!$ is put into effect in such a way that an error (unintended state) occurs with a failure of $W\!$ as a consequence. A usage profile provides information about probabilities of actual usage in relation to the corresponding parameter conditions

\subsection{Failure probability for inseqware usage}
In some cases authorities responsible for operating complex systems require of their engineers that for inseqware $W\!$ serving within the control of a particular system $S\!$ a probability of failure of $W\!$ (and caused by a fault in $W\!$) is produced, or at least a comfortable upper bound of that probability is given, before a usage decision for $W\!$ can be taken.

Whether such a probability can be adequately defined with a definition based on some form of probability theory for system behavior is another matter. A useful simplifying assumption is that probabilities of current and future actual usage under a full variety of parameter conditions can be estimated. That leads to a so-called usage profile for $W\!$ in the context of $S\!$. Now in order to speak of a failure risk in connection with $W\!$, one needs to obtain information about the probability that effectuation of $W\!$ leads to failure under specific parameter conditions. At this stage, however, the ontology of $W\!$ starts to matter. If $W\!$ is conceived of as a mathematical entity the probability of failure of $W\!$, again given particular parameter conditions, is either zero or one. In order to see this it suffices to notice that if an error occurs during effectuation, given the mentioned parameter setting, that must happen by necessity, (given that same parameter setting), and conversely if it does not occur it could not have occurred (for that same parameter setting).

At the other end of the spectrum of existence in an inseqware ontology, inseqware $W\!$ is known through the mediation of a particular physical representation $W_p\!$ together with an abstraction $A\!$  of that representation's history of coming about. In that case forthcoming experimentation may be required to say anything meaningful about the probability of failure (given the mentioned parameter conditions). This second view is also problematic. Because it may carry with it too much information an abstraction of that information must be applied out in advance of further analysis. To grasp the impact of such uncertainties the following questions about the history of $W_p$ may be considered: 
\begin{enumerate}
\item Suppose that one of the engineers (say $E$)  who was involved in the production of $W\!$ had malicious intentions, as judged from the perspective of the owner/operator of $S\!$. Is that state of affairs a property of $W_p\!$ even if that state of affairs went by unnoticed and as a consequence it has not been recorded in $A$? 
\item Suppose that $E$'s malicious intentions were noticed though  by accident it has been forgotten to include that information in the documentation files when $A$ was collected. Is it a property of $W_p$ impacting failure probability estimates?
\item And if such information was obtained but has been subsequently and intentionally deleted by malicious quality control staff, and so on? 
\item Can one accept that a judgement about failure probabilities for a given technical item (specimen of inseqware), about which  information regarding its current state is known, is yet  highly dependent on fragmentary historical data.
\end{enumerate}
Answers to one or more of these questions constitute judgements that might impact a risk analysis concerning the failures that may be caused by faults in $W\!$.

\subsection{Subjective probabilities may help}
It is a plausible step to start thinking in terms of subjective probabilities depending on the path along which information about $W_p$ has been revealed in successive stages to an observer in charge of the pertinent  risk analysis. A Bayesian approach is conceivable where for instance prior odds attribute (i) a low, say $10^{-6}$, probability to a an engineer being malicious, and (ii) a high probability, say $1-10^{-6}$, to a malicious engineer succeeding in covertly inserting a fault in $W\!$ which will lead to failure under realistic parameter settings, a failure which, moreover, will cause a significant malfunction of $S\!$.

The prior odds may be estimated with more confidence on the basis of historical data of comparable cases. Working with subjective reasoning still has to be combined with default reasoning concerning a range of problems that might have occurred but about which no data are available. For instance an engineer may have been ill or the engineering team may have suffered from internal conflicts.

\subsection{Fictional engineers and fictional production processes}
Instead of worrying about which information about $W_p$'s coming into existence has been preserved, a simpler view might be as follows: a story is told by a fictional narrator about how $W\!$ was produced, 
this narrative being phrased as a fictional episode rather than as a statement of historic fact. 

The description of the inseqware production process contains so many particular details that it is a useless attempt to perform statistics on a large collection of logfiles from observed factual historic production processes of comparable inseqware. The probability that other processes that satisfy the same specification (process description) are found is almost zero. The particular production process at hand is best understood as one among a family of alternative and presumably equally fictional production process descriptions about which uniform methods of quality assessment and risk analysis must be developed and applied. 

Then the fictional production process, thus specified by way of storytelling, is equipped with a risk analysis. After a suitable comparative quantification of various items occurring in that risk analysis a figure is found that is referred to as an (upper bound) of a probability of the occurrence of a fault (in  the fictional end product 
$W_f\!$ of said fictional production process).%
\footnote{The definition of what constitutes a fault in a fictional inseqware $W\!$  requires careful attention, and occurrence of a failure must be connected with fictional parameter conditions rather than with real ones.}
Subsequently the match between the story line and what is known about the actual production process of $W\!$ is assessed and if that match is found convincing the same probability is now used as a judgement (estimate) about the failure probability of $W\!$ itself.

One might question the role of fiction in this description of the risk assessment process because all that seems to
 matter is abstracting from a given reality. The problem with the given reality, however, is that it is unknown what is unknown about it and dealing with such uncertainties is difficult. The computer science answer to such complications is to make a simplified model and to formalize that model. Having done so the risk analysis is performed in the model and it is turned into applied mathematics and logic. That approach clearly fails if people need to be modeled. If human beings feature inside a mathematical model that model is better understood as a chapter in fiction than as a mathematical formalization. Fictional characters make an appeal to a reader's ability to fill in informational gaps by 
 taking on board plausible additional (and fictional) assumptions in a way that elements of a logical formalization cannot do.

\subsection{Fictional engineering in practice} 
These considerations emerged from working in the context of a contracted project on software quality assessment of which the details cannot yet be disclosed at this moment. The preliminary conclusion that I have drawn is that I got involved in a case where quality assurance for safety critical inseqware invites one to think in terms of fictional 
engineers, fictional production processes, and of course fictional software resulting from such processes. To the fictional production process a form of quantified qualitative reasoning, (fictionally) simulating the Bayesian method of updating (fictional) probabilities in successive stages, is applied with the aim of producing a fictional probability figure. That figure (pertaining to 
$W_f$)can subsequently be used as a substitute of a ``real probability estimate" (that is an estimate relevant for 
$W_p$) under the assumption that the fictional process was considered to be sufficiently close to the observed process.

Based on a theory of fictional accounts of inseqware production, one may (hypothetically) acquire conjectural abilities (see \cite{BDV2011b}) concerning the reliable, or at least reproducible, production of (fictional) failure probabilities. Filling in a questionnaire and computing a cumulative score may be an appropriate competence underlying that ability. Designing the questionnaire and scoring methods constitute research problems for any given theory of fictional inseqware engineering processes. In any case a theory of fictional inseqware development processes must be made so rich in detail that an appropriate questionnaire with scoring technique can be found.

Providing accounts of fictional  inseqware development processes is a matter that has not been worked out in any detail to the best of my knowledge. Apart from a need to take a theory of fiction on board, many matters need to be contemplated, such as for instance the status of promises (see \cite{Burgess2005,Burgess2007} in a world of fiction. (Fictional) promises enter the picture for instance if the quality of work is to be connected with explicit professional standards and ethical codes of conduct constraining the activity of formally qualified and certified (fictional) personnel.

\section{Concluding remarks}
\label{Conclusions}
Starting with a proposal on Dutch terminology and its translation to English, fiction science has been featured as a component of the academic (scientific, scholarly) counterpart of informatics. Fiction seems to be very distant from computer science, but fiction enters through the front door of computer science if modeling production processes for safety critical systems takes human actors into account as abstract entities. 
Adequate forms of default reasoning that apply to such cases can perhaps be best understood as methods to assess fictional narratives. Fiction science may be needed to provide a methodological basis for such reasoning patterns. Assessment of analogy between fiction and observed fact provides the inference mechanism for the informal logic governing the application of this kind of (applied) fiction.

Besides supporting applications in conventional systems engineering fiction science emerges as a (potential) host for computer gaming, a field of ever increasing impact. Having elevated fiction and its (as yet fictional) academic container fiction science (FS)
to these heights relative to other domains of computing,  the equations  IY = CS + DS + FS and INF = P(IY) = P(CS + DS + FS) acquire some  plausibility.

\end{document}